\documentclass[fleqn,10pt]{wlscirep}
\title{Subwavelength Interferometric Control of Absorption in Three-port Acoustic Network}

\author[1,*]{O.~Richoux}
\author[1]{V.~Achilleos}
\author[1]{G.~Theocharis}
\author[2]{I.~Brouzos}
\affil[1]{Le Mans University, LAUM UMR CNRS 6613, Av. O. Messiaen, 72085 Le Mans, France}	
\affil[2]{Department of Physics, University of Athens, 15771 Athens, Greece}

\affil[*]{Olivier.Richoux@univ-lemans.fr}

\usepackage{graphicx}
\usepackage{amsmath}

\begin{abstract}
Utilizing the effect of losses, we show that symmetric 3-port devices exhibit
coherent perfect absorption of waves and we provide the corresponding conditions on the reflection and transmission
coefficients. Infinite combinations of asymmetric inputs with different amplitudes and phase at each port as well as a 
completely symmetric input, are found to be perfectly absorbed.
To illustrate the above we study an acoustic 3-port network operating in a subwavelength frequency both theoretically and experimentally. In addition we show how the output from a 3-port network is altered, when conditions
of perfect absorption are met	 but the input waves phase and amplitude vary. In that regard, we propose optimized structures 
which feature both perfect absorption and perfect transmission at the same frequency by tuning the amplitudes and 
phases of the input waves. 
\end{abstract}
\begin{document}

\flushbottom
\maketitle

\thispagestyle{empty}

\section*{Introduction}

The absorption of wave energy is a phenomenon which underlies many applications in acoustics and photonics including 
molecular sensing~\cite{Cubukcu,Lee}, photodetection~\cite{Yu}, and sound proofing~\cite{Mei,Yang1}. By exploring wave interference, 
the absorption can become much more efficient or even complete.
Indeed, coherent perfect absorbers have been extensively studied the last years in different photonic~\cite{Ramezani16,Chong,Longhi,Sun,Zanotto,Zanotto16,Zhu,Baldacci} and acoustic~\cite{Merkel,Meng,Li} structures.
However, the majority of these works deals with two-port systems. 
Especially in acoustics, the phenomenon of perfect absorption has attracted great attention the 
last years due to its direct applications to numerous noise reduction problems. 
Many solutions have been proposed in the low frequency regime based on subwavelength metamaterial designs,
by critically coupling~\cite{Xu00} resonant scatterers to the waveguide  i.e. by balancing the energy leackage and the internal
losses of the resonators. Such studies include the use of membranes~\cite{Mei,Ma,Wei,Romerosr,Meng,Long}, quarter-wavelenght structures~\cite{yongli2016,Cai,Jiang}, 
the concept of slow sound~\cite{Groby1,Groby2,Yang} and Helmholtz resonators both in the linear~\cite{Merkel,Romero16,Noe} and nonlinear~\cite{Achilleos16} regimes.

Muti-port devices have a great use in applications such as radio frequency (RF) systems and signal/ information processing. For example in microwave physics, devices such as power dividers, circulators, filters couplers and multiplexer are commonly used in many RF systems~\cite{Gil,Shang}. On the other hand, complex photonic circuits have been studied and designed for optical signal
processing or computing in integrated optics~\cite{Parinaz,Pan}. 
From theoretical perspective, the study of complex multi-port networks is also a vibrant area of research attracting
considerable attention in different fields, including plasmonics~\cite{Veronis,Feigenbaum2007,Feigenbaum2010,Feigenbaum}, quantum transport~\cite{Roger,Akguc} and acoustics~\cite{Moleron,Zhang22,Wei22,Peng}.

Although Coherent Perfect Absorption (CPA) has been thoroughly studied both experimentally and theoretically in two-port systems, 
it is only recently that multi-port devices have been proposed as perfect absorbers. In particular in Ref.~\cite{Zhang}, in an 
asymmetric three-port acoustic  device it was found that acoustic energy could be channeled from one port to another 
using a phase mismatch of the input. Additionally adding an additional branch in a standard PT-symmetric
electromagnetic waveguide~\cite{Fu,Ramezani2}, it was shown that it is possible to achieve asymmetric output from this branch when the system is excited
from either the loss or the gain side of the main waveguide.

Here, using subwavelength resonators and interferometric control of absorption we illustrate, both theoretically and experimentally, 
an acoustic perfect absorbing network that operates at different wavelengths, different intensities and relative phases of the input waves. It is found that, in great contrast to the two-port case, there is an infinity of input wave combinations that can be completely absorbed, when the device satisfies CPA conditions
Moreover, we propose optimized 3-port networks which operate both as perfect absorbers and 
Coherent Perfect Transmitters (CPT), operating at the same frequency. We show how these systems undergo a transition
from CPA to CPT by just tuning the phase and/or the amplitude of the input waves. A high contrast of
output to input power ratio is obtained, due to the use of point-like acoustic scatterers . Therefore we believe that
utilizing the proposed subwavelength, interferometric control of absorption in multi-port devices and exploring further the role of symmetries could have a strong impact in the field of compact wave devices.

In  Section 1, we study the general case of a three port, single-mode, scattering system and obtain the necessary conditions
for achieving CPA. The conditions depend both on the device scattering properties and the coherence of the three incoming waves.
 We show both theoretically and experimentally that an acoustical network composed  by three waveguides side-loaded with Helmholtz resonators, can be tuned in order to satisfy these conditions, and CPA is achieved for different type of inputs ( asymmetric and symmetric) in the subwavelength frequency regime.
In Section 2,  we study the interferometric control of the network using different input vectors when the system is in the CPA
configuration.  We obtain the necessary conditions under which a 3-port network exhibits both CPA and CPT. Using an optimization process, we propose realistic networks satisfying these conditions and verify our results using finite elements 3D simulations.
\section{Symmetric 3-port system}
\begin{figure}[htb]
\includegraphics[width=14cm]{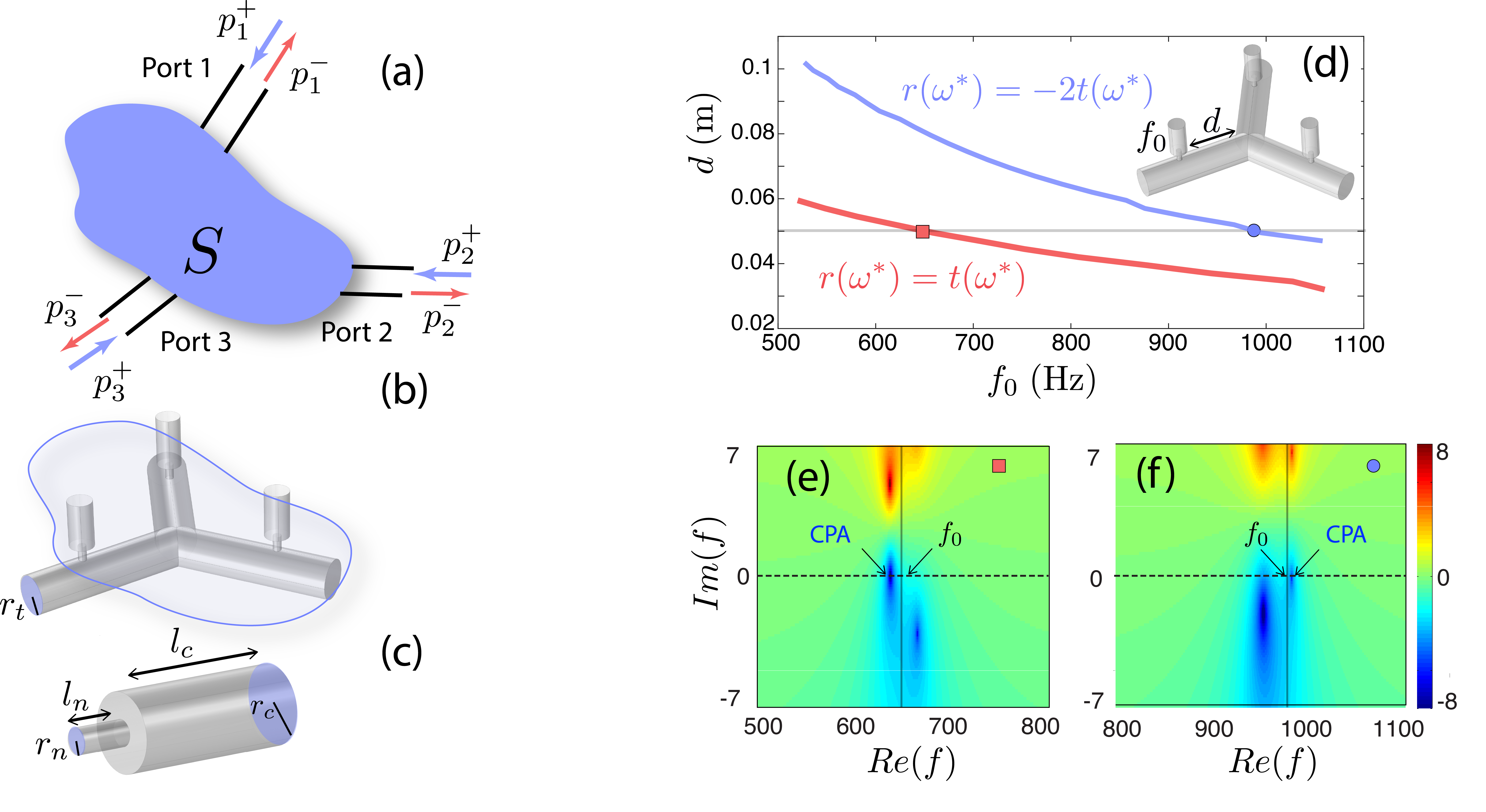}
\caption{(a) A schematic illustration of  a general three port system with incoming and outgoing waves at each port.
The scattering matrix $S$ of the system is assumed to be symmetric, at some frequency range, even if the geometry of the
device may be not. (b) The symmetric network studied here (not in scale), which is composed by 3 identical cylindrical 
waveguides with radius $r_t=2.5\times 10^{-2}$~m, assembled by a Y-shape connection sideloaded with HRs at a distance $d$.
(c) Details of the HRs  composed by a neck with  length $\ell_n=2\times 10^{-2}$~m, a radius $r_n=0.45\times 10^{-2}$~m branched to a cylindrical cavity with radius $r_c=1.5\times 10^{-2}$~m  and  varying length $\ell_c$
which is used in order to tune the resonance frequency $f_0$.
(d) The upper [lower] curve depicts the symmetric [asymmetric] CPA solutions defined the by Eq.~(\ref{cond1}) [Eq.~(\ref{cond2})], for varying resonant frequency $f_0$ and distance $d$.
The horizontal line corresponds to the configurations with  $d=0.05$~m.
(e),(f) The determinant $|{\rm det}(S)|$ in the complex frequency plane for the configurations
corresponding to the two cases of panel (d) with a (red) square and a (blue) circle respectively. 
}
\label{sketch_gen} 
\end{figure}

\subsection{Scattering properties}

In this work, we study reciprocal and symmetrical 3-port acoustic networks and we focus our analysis
 on the corresponding scattering matrix which is given by the following equation
\begin{equation} 
\label{Eq-Scattering}
\begin{pmatrix}
p_1^- \\
p_2^- \\
p_3^-
\end{pmatrix}
=\begin{pmatrix}
r & t & t\\
t & r & t  \\
t & t & r \\
\end{pmatrix}
\begin{pmatrix}
p_1^+ \\
p_2^+ \\
p_3^+
\end{pmatrix} 
= S
\begin{pmatrix}
p_1^+ \\
p_2^+ \\
p_3^+
\end{pmatrix} .
\end{equation}
In Eq.~(\ref{Eq-Scattering}) the vectors  $(p_1^+,p_2^+,p_3^+)^T\equiv|\psi_{\rm in}\rangle$ and  $(p_1^-,p_2^-,p_3^-)^T\equiv|\psi_{\rm out}\rangle$ describe  the incoming and 
outgoing waves respectively as shown in Fig.~\ref{sketch_gen}(a). The frequency dependent coefficients $r$ and $t$, correspond to the reflection and transmission when only one port is excited as explained in details in the Appendix. Note that the S-matrix
as given by Eq.~(\ref{Eq-Scattering})  is a \textit{symmetric circulant} matrix which stemms both from the geometrical symmetry
of the 3-port network and from the fact that we consider a reciprocal system.

Furthermore the eigenvalue problem associated with the S-matrix is given by
\begin{eqnarray}
{\rm det} (S-\lambda I)=0,
\label{eigenprobelm}
\end{eqnarray}
where the eigenvalues are found to be:
\begin{equation}
\lambda_0 = r+2t, \mbox{ and } \lambda_1 = \lambda_2 = r-t, 
\label{eigvalues}
\end{equation}
and the corresponding orthonormal eigenvectors are given by:
\begin{eqnarray}
|u_0\rangle =\frac{1}{\sqrt{3}}(1,1,1)^T \label{u1},\quad |u_1\rangle =\frac{1}{\sqrt{3}}(1, e^{2i\pi/3}, e^{-2i\pi/3})^T\label{u2}
, \quad |u_2\rangle =	\frac{1}{\sqrt{3}}(1, e^{-2i\pi/3}, e^{2i\pi/3})^T.\label{u3}
\end{eqnarray}
It is interesting to note here that, the eigenvectors of a circulant matrix are always the same and are independent of 
both the physical system (particular form of $r$ and $t$) and the frequency.

The above eigenanalysis is useful for the study of CPA since if an eigenvalue $\lambda_i$ of $S$ is found to be zero,
then using the corresponding eigenvector $|u_i\rangle$ as input in Eq.~(\ref{Eq-Scattering}) will result in zero
output, thus perfect absorption.
To quantify the absorption efficiency of the 3-port network below we use the ratio of total output to input power defined as
\begin{equation}
\Theta = \frac{\sum_i |p^-_i|^2}{\sum_i |p_i^+|^2}=\frac{\sum_i \lambda_{i-1}^2 |c_{i-1}|^2}{\sum_i|c_{i-1}|^2 } \quad i=1,2,3.
\label{eq:Theta}
\end{equation}
In the last expression we have used the fact that, since the eigenvectors $|u_i\rangle$  form a complete basis, we may
write any input vector as a sum of  $|u_i\rangle$,  i.e.  $|\psi_{\rm in}\rangle=\sum_i c_i|u_i\rangle$.

\subsection{Coherent Perfect Absorption}

According to the eigenanalysis, a 3-port network exhibits CPA if any of the eigenvalues given by Eqs.~(\ref{eigvalues}) 
becomes zero at some particular frequency $f^*$.
We first consider the case of  $\lambda_0=0$, which leads to the following condition for the scattering coefficients
\begin{equation}
r=-2 t\equiv r_s.
\label{cond1}
\end{equation}
If Eq.~(\ref{cond1}) is satisfied, then a symmetric input of the form $|u_0\rangle$ [Eq.~(\ref{u1})] will be completely absorbed. Below we refer to the combination of Eq.~(\ref{cond1}) and an input  $|u_0\rangle$  as {\it symmetric} CPA.

On the other hand, by setting the degenerate eigenvalues $\lambda_{1,2}=0 $ we obtain a different CPA condition 
\begin{equation}
r=t\equiv r_a.
\label{cond2}
\end{equation}
When Eq.~(\ref{cond2}) is satisfied, an input wave in the form of the asymmetric vectors $|u_1\rangle$ or $|u_2\rangle$
is completely absorbed. Moreover, since the system is linear, any input in the form of 
$|\psi_{\rm in}\rangle=\alpha |u_1\rangle+\beta |u_2\rangle$ (with  $|\alpha|^2+|\beta|^2=1$) 
will also be completely absorbed.
This is in great contrast with two port systems where CPA is achieved  either by in-phase or out-of phase incoming waves.
Here, the additional port acts as a ``control'' port which, whenever Eq.~(\ref{cond2}) is satisfied, 
is able to lead the 3-port network to CPA.
We further on refer to this combination of condition (\ref{cond2}) and the inputs as {\it asymmetric} CPAs.

The conditions to achieve CPA given by Eqs.~(\ref{cond1}) and (\ref{cond2}) require that the transmission and reflection coefficients
are specifically tuned. A popular way to achieve this is to use Fano resonance phenomena by employing resonant scatterers, 
especially regarding subwavelength manipulation of waves. In acoustics the most prominent example of such a scatterer is the Helmholtz resonator (HR), which has been intensively studied in the context of wave absorption and acoustic metamaterials.
The particular system under consideration in this work is schematically shown in Fig.~\ref{sketch_gen}(b) and (c). 
We consider all HRs to have identical resonance frequencies $f_0$ and to be placed at the same distance $d$ from the center of the device. In the low frequency regime, assuming only the propagation
of the plane mode of the waveguides and approximating the HRs as point scatterers  we can calculate $t$ and $r$  using the standard transfer matrix method. The analytic calculations of $t$ and $r$ are detailed in the section Method.

In order to obtain the configurations which exhibit CPA, we calculate $r$ and $t$ by scanning the parametric space
($f_0$, $d$) and monitor when the conditions of Eqs.~(\ref{cond1}) and (\ref{cond2}) are 
satisfied. The rest of the parameters are fixed to the experimental values given in the caption of Fig.~\ref{sketch_gen}. The results are shown in Fig.~\ref{sketch_gen}(d) where the red (lower) curve  corresponds to the asymmetric
and the blue (upper) line depicts  the symmetric CPA.
The occurrence of CPA is commonly illustrated in the complex frequency plane of the determinant of the scattering matrix $S$.
In such a representation CPA is associated with a zero of the determinant of $S$ which is located at the real frequency axis, 
and its location indicates the operating frequency $f^*$. 
In Fig.~\ref{sketch_gen}(e) and (f) , we plot the ${\log|\rm det}\;  S|$ for two different configurations which satisfy 
Eqs.~(\ref{cond1}) and (\ref{cond2}) respectively. It is directly shown that in both cases, one zero of the determinant
is located on the real axis.
Note that, as indicated in Fig.~\ref{sketch_gen}(e) and (f), the operating frequency for the CPA devices
is near (but not the same) to the resonant frequency of the HRs due to the interaction of the resonances  of each HR
through the waveguides.

\subsection{Experimental observation of acoustic CPA}
\begin{figure}[t]
\includegraphics[width=15cm]{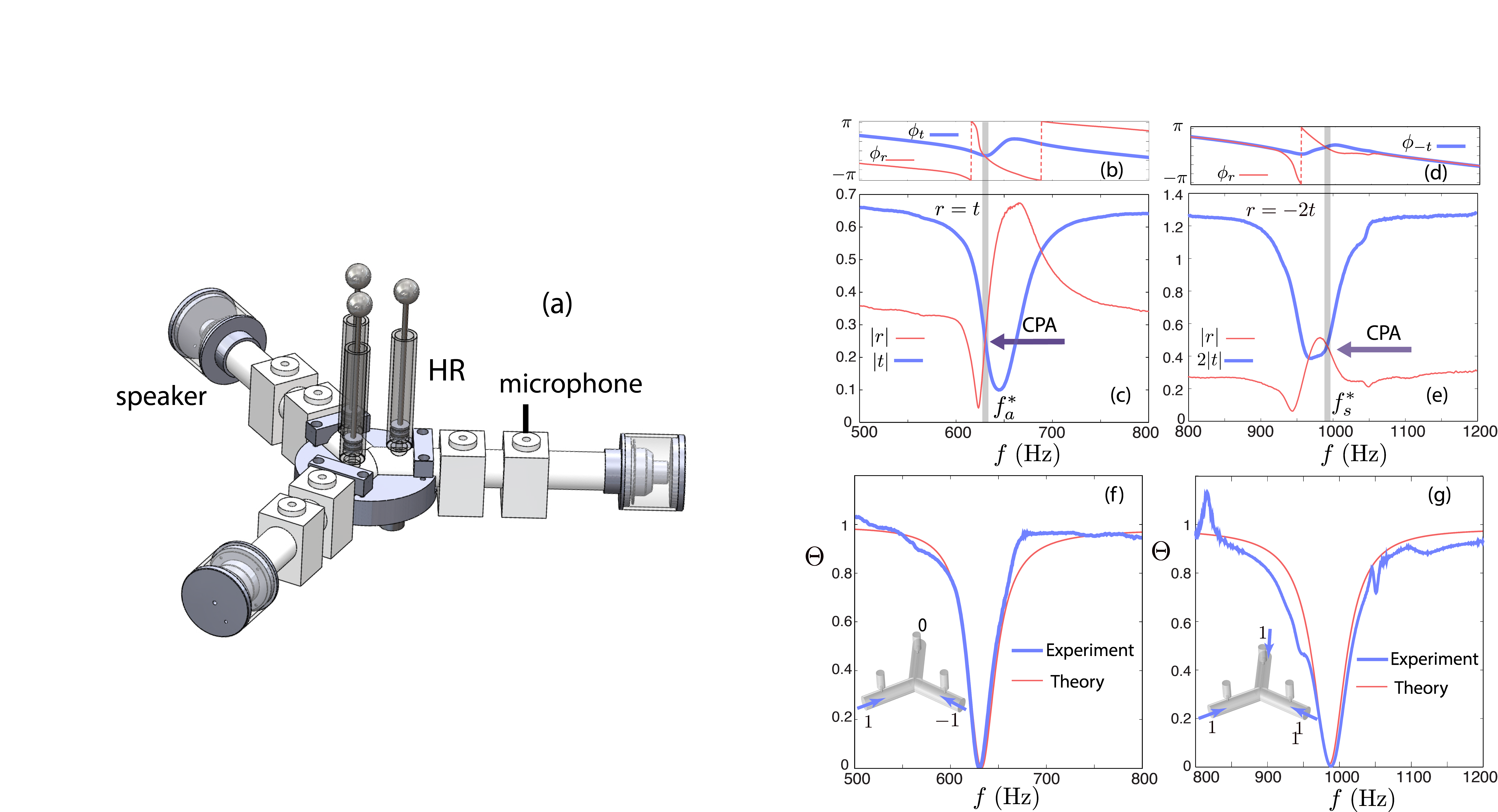}
\caption{(a) View of the experimental device. (b), (c) The phase and absolute value of the experimentally measured reflection and transmission coefficients
$t$ and $r$, as a function of frequency, for the configuration which exhibits an asymmetric
CPA at $f_a^*=630$~Hz. CPA is ensured since both the real and the imaginary part of $r$ and $t$ are equal for this 
frequency (vertical gray line). (d), (e) The phase and absolute value of the experimentally measured reflection and
transmission coefficients $r$ and $-2t$, as a function of frequency, for the configuration which exhibits a symmetric
CPA at $f_s^*=988$~Hz. CPA appears when the two the curves of both the real and imaginary part become equal,
indicated by the vertical gray line. (f) and (g) The output to input power ratio $\Theta$ as a function of frequency for the asymmetric and 
the symmetric CPA respectively. The thick (thin) line corresponds to the experimental (theoretical) measurement.
The insets depict the eigenvector used in order to obtain the curves in each case.}
\label{fig3} 
\end{figure}
In this Section, we experimentally verify the analytical results of Fig.~\ref{sketch_gen}(d)  in an acoustic network 
with  $d=0.05$~m. We use two sets of HRs with different resonant frequencies $f_0$ corresponding to the asymmetric
 and symmetric CPA as indicated respectively by the square and the circle in Fig.~\ref{sketch_gen}(d).
In our experiments, the 3-ports network is driven with plane waves produced by loudspeakers (AURA
NS3-193-8A 3 inch)  placed at the end of each port (see Fig.~\ref{fig3}(a)). The system is considered to 
be symmetric as long as  the driving frequencies are below the first cut-of frequency of the waveguides such that 
no higher modes are propagating.
To determine experimentally the reflection and transmission coefficients, a pair of microphones (1/2 inch B\&K)
connected to each waveguide is employed, allowing the measurement of both
the forward $p_i^{+}$ and backward $p_i^{-}$ waves in each waveguide~\cite{Song,Muehleisen}.  Using the definition of the scattering matrix Eq.~(\ref{Eq-Scattering}) and the measured values of the incoming and outgoing waves, we can directly obtain $t$ and $r$. 

We perform measurements for a range of driving frequencies between $400$~Hz to $1200$~Hz using a sweep sine function,
and the results are shown in Figure~\ref{fig3}. Panels (b) and (c) correspond to the asymmetric CPA. 
For this setting it is directly shown
that at  frequency $f_a^*=630$~Hz the required condition for the asymmetric CPA $r=t$  is fulfilled as indicated
by the vertical gray line. Similarly, for the configuration  corresponding to the symmetric case,  as shown in Fig.~\ref{fig3}(d) and (e)  
the condition $r=-2t$, is also fulfilled at the CPA frequency $f_s^*=988$~Hz (vertical gray line).

By measuring the complex coefficients $r$ and $t$ we have experimentally determined the scattering matrix of the 3-port 
network for the two different configurations  for the frequency range of interest.
We can thus now quantify the ability of our setup to completely absorb an input wave at the CPA frequency,
by considering an asymmetric input vector $(p_1^+,p_2^+,p_3^+)^T=(1,-1,0)^T$ and the 
experimentally obtained scattering matrix in Eq.~(\ref{Eq-Scattering}). Then from the 
corresponding output we may calculate $\Theta$ by its definition from Eq.~(\ref{eq:Theta}). 
The result is shown in Fig.~\ref{fig3}(f) with the thick solid line, where at the CPA frequency $f_a^*$ 
we obtain an almost perfect absorption with $\Theta\approx 5\times 10^{-4}$. The experimental result is in a good agreement
with the theoretical prediction [thin solid line in Fig.~\ref{fig3} (f)] calculated using the transfer
matrix method.

We perform the same analysis for the network with the symmetric CPA.
In particular, considering an input of the form $u=(1,1,1)^T$ [inset of Fig.~\ref{fig3}(g)]
and using the experimental values of $r$ and $t$ [see Fig.~\ref{fig3}(d)-(e)] we obtain $\Theta$  for a frequency range between $800$~Hz to $1200$~Hz. 
The result is shown with the thick solid line in Fig.~\ref{fig3} (g), and in this case the system reaches a value of 
$\Theta\approx 10^{-3}$ at $f_s^*$. The peaks appearing on the experimental curve are due to
small structural imperfections of our device and limitation of the two microphone method, in this frequency range.
The comparison with the theoretical result illustrated with the thin line Fig.~\ref{fig3} (g) shows a good agreement between the two, close to the CPA frequency. 
\section{Interferometric control of the 3-port network}
Up to this point we have shown that when a 3-port network is tuned to satisfy either 
Eq.~(\ref{cond1}) or (\ref{cond2}) then it completely absorbs certain combinations of input waves
from each port, corresponding to the $S$ matrix eigenvectors (or their linear combinations).
Here we study the behavior of such a tuned network, varying the input waves such that
we deviate from the perfectly absorbed eigenvectors and the device transmits some amount of energy.
In fact our goal is to identify the conditions under which a 3-port network acts both as 
perfect absorber ($\Theta=0$) and a perfect transmitter  ($\Theta=1$) at the same frequency.

According to Eq.~(\ref{eq:Theta}), an input vector in the form of an eigenvector of $S$, i.e.
$|\psi_{\rm in}\rangle=|u_i\rangle$ will result in $\Theta=\lambda_i^2$ as output to input ratio.
If for the same parameter values and at the same frequency there is an eigenvalue $\lambda_i=0$
and one with $\lambda_{j\ne i}=1$ then the device exhibits both CPA and CPT.
This constrain defines a pre-described value for the coefficients $r$ and $t$ which is
different depending whether Eq.~(\ref{cond1}) or (\ref{cond2}) is satisfied.
%

\subsection{From asymmetric CPA to symmetric CPT}

We first consider the case when Eq.~(\ref{cond1}) is satisfied and thus $\lambda_{1,2}=0$ leading to CPA with 
an asymmetric input $|\psi_{\rm in}\rangle~=~\alpha |u_1\rangle~+~\beta |u_2\rangle$. Demanding also
 $\lambda_0=1$ leads to $r_a=t=1/3$ [see Eq.~( \ref{eigvalues})] and for the same 3-port network
a symmetric input  $|\psi_{\rm in}\rangle =|u_{0}\rangle$ results in CPT.
Let's note that, for our experimental device, as illustrated in Fig.~\ref{fig3}(b), the measured value of the reflection 
coefficient is $|r_a^{\rm exp}|=0.25$ and thus CPT cannot be reached.

In order to check if a realistic device that satisfying both the conditions for CPA and CPT can be found, we use an optimization algorithm. 
We first choose a prescribed operating frequency close to asymmetric CPA
experimental value $f_a^*=630$~Hz and then we optimize the geometrical characteristics of the device
to achieve the prescribed condition $r_a=t=1/3$ at $f_a^*$ ($r$ and $t$ are calculated using the analytical expressions 
from the transfer matrix method). The optimization method converges up to $99\%$ with  $r^{\rm op}_a=0.33$ and the resulting  geometrical parameters of the optimized device are given in Ref.~\cite{pars1}.

\begin{figure}[t]
\includegraphics[width=8.5cm]{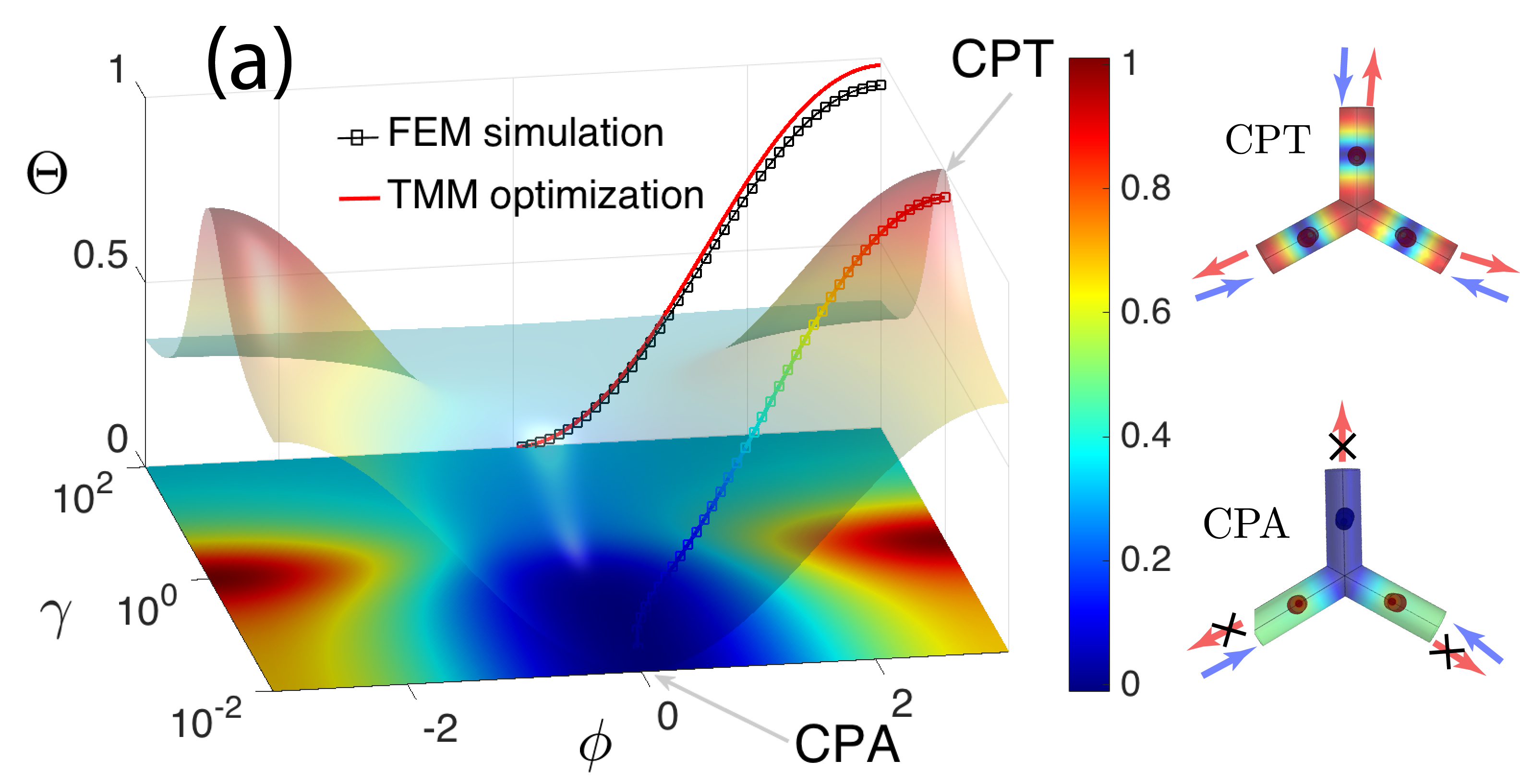}
\includegraphics[width=8.5cm]{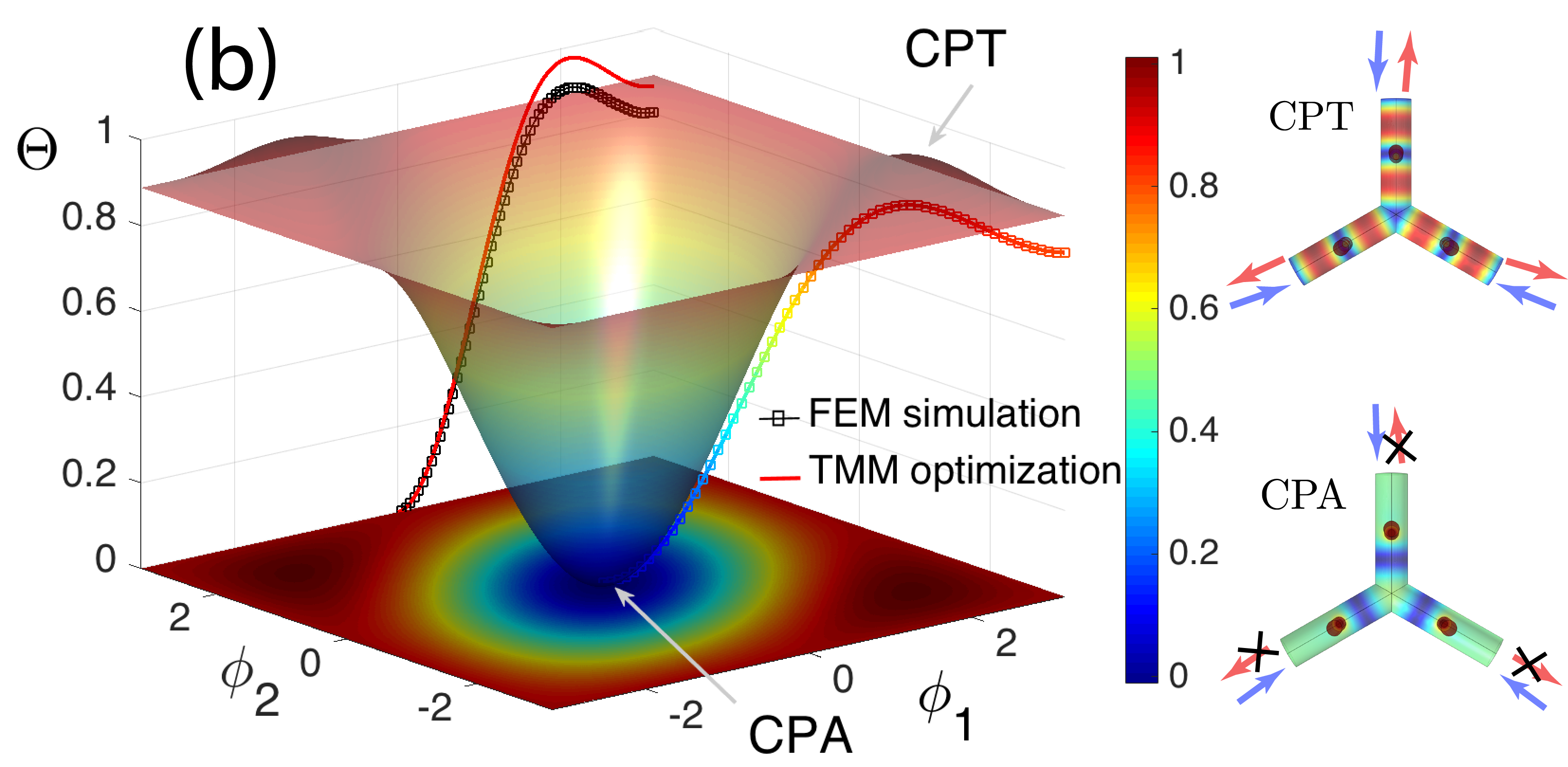}
\caption{(a) The output to input power ratio $\Theta$ as a function of  $\gamma$ and $\phi$ using the
input vector $u_{I_a}$, when the system is in the optimal configuration for asymmetric CPA
with $r_a^{\rm op}=0.33$, as given by Eq.~(\ref{eq:Theta_asym}). 
At the $(\phi,\Theta)$ plane we show the result obtained with the  FEM simulation (squares) and with 
the result of  Eq.~(\ref{eq:Theta_asym}) (solid line), along the trajectory $\gamma=\phi/\pi$ with $\gamma\in[0,1]$.
On the righthand side we plot the field distribution of the absolute pressure in the network at $\gamma=\phi=0$ (CPA)
and at $\gamma=\phi/\pi=1$ (CPT) as obtained by the FEM simulation.
(b) The output to input power ratio $\Theta$ as a function of  $\phi_1$ and $\phi_2$ using the
input vector $u_{I_s}$ with $\gamma_{1,2}=1$, when the system is in the optimal configuration for symmetric CPA
with at  $r_s^{\rm op}=0.657$, as given by Eq.~(\ref{eq:Theta_asym}). 
In the $(\phi_1,\Theta)$ plane we show the result obtained with the  FEM simulation (squares) and with 
the result of  Eq.~(\ref{eq:Theta_asym}) (solid line), along the trajectory $\phi_1=-\phi_2$ with $\phi_1\in[0,\pi]$.
On the righthand side we plot the field distribution of the absolute pressure in the network at $\phi_{1,2}=0$ (CPA)
and at $\phi_1=-\phi_2=2\pi/3$ (CPT) as obtained by the FEM simulation.}
\label{fig5} 
\end{figure}

To study the transition from CPA to CPT for the optimized network, we calculate $\Theta$  using an input vector of the form
\begin{equation}
u_{I_a}=(1,- e^{i\phi},\gamma)^T.
\label{uIa}
\end{equation}
 Here, $\gamma$ is a real number characterizing  the ratio between the amplitudes of the incoming waves and  $\phi$ denotes 
the phase difference from the CPA eigenvector (defined by $\gamma=0$ and $\phi=0$).
Using Eqs.~(\ref{Eq-Scattering}), (\ref{cond2}) and~(\ref{eq:Theta}) we find that 
\begin{equation}
\Theta(\gamma,\phi)= 3| r_a |^2 \Big(1+\frac{2\big(\gamma-\cos\phi(1+\gamma)\big) }{2+\gamma^2}\Big).
\label{eq:Theta_asym}
\end{equation}
In Fig.~\ref{fig5}(a) we plot the parameter $\Theta$ as a function of $\gamma$ and $\phi$. We identify the
CPA point at $\gamma=\phi=0$ and the CPT points at $\gamma=\pm \phi/\pi=1$. From 
Eq.~(\ref{eq:Theta_asym}) and as illustrated in Fig.~\ref{fig5}(a) the CPA and CPT
points, are the extrema of this function. As such, small deviations around these points
will lead to a {\it quadratic sensitivity} of the two phenomena. The solid line in the $(\phi,\Theta)$ plane shows $\Theta$
for a particular trajectory defined by  $\gamma=\phi/\pi$ where $\gamma\in[0,1]$, 
which starts at CPA ($\gamma=0$) and ends at CPT ($\gamma=1$).

In addition, to verify the device efficiency of the optimized device in the full 3D space, we use a 3D FEM simulations and study the corresponding geometry for the trajectory $\gamma=\phi/\pi$ with $u_{I_a}$. The result of the FEM simulation
is shown with the squares both on top of the 3D contour and on the $(\phi,\Theta)$ plane.
There is a good agreement between the two results, verifying our 
model using the transfer matrix method. The 3D network is able to achieve an almost total absorption with an output ratio
$\Theta\approx10^{-3}$, and an almost perfect transmission with $\Theta = 0.93$. 

\subsection{From symmetric CPA to asymmetric CPT}
For the case where  Eq.~(\ref{cond2}) is satisfied and thus  $\lambda_{0}=0$, the 
network exhibits CPA for symmetric inputs of the form  $|\psi_{\rm  in}\rangle~=~|u_{0}\rangle$. 
Here the additional condition for CPT  is  $\lambda_{1,2}=1$ which leads to  $r_s=-2t=2/3$. Then the device will
also completely transmit asymmetric inputs of the form $|\psi_{\rm in}\rangle~=~\alpha~|u_1\rangle~+~\beta~|u_2\rangle$.
For this case our experimental setup featuring a value $|r_s^{\rm exp}|=0.52$ is not able to reach CPT.

Thus, an optimization scheme is again used to obtain a device that achieves both CPA and asymmetric  CPT.
We choose to optimize at a frequency close to the experimental symmetric CPA frequency $f_s^*=988$~Hz.
The optimization method converges up to $98.5\%$ corresponding to $r^{\rm op}_a=0.657$ and the parameters
of the optimized device are given in Ref.\cite{pars2}.

In a similar way as in the asymmetric case, we study the dependence of $\Theta$ as a function of two parameters, using the input 
vector
\begin{equation}
u_{I_s}=(1,e^{i \phi_1},e^{i\phi_2})^T,
\label{uIs}
\end{equation}
where $\phi_{1,2}$ quantify the phase differences from the CPA eigenvector. 
Using Eqs.~(\ref{Eq-Scattering}) and (\ref{eq:Theta}) we obtain that
\begin{align}
&\Theta(\phi_{1,2}) = \frac{3}{2}\vert r_s \vert^2 \left(1-\frac{\cos\phi_1+\cos\phi_2+\cos(\phi_1-\phi_2)}{3}\right).
\label{Theta_symmetric}
\end{align}
The dependence of $\Theta$ on the two phase differences is plotted in Fig.~\ref{fig5}(b). We observe that the maximum of $\Theta$
as we discern from the CPA eigenvector, is obtained when the two phase differences satisfy $\phi_1=-\phi_2=2\pi/3$,
i.e. when $|u_{I_s}\rangle=|u_{1,2}\rangle$. In this case also, as illustrated in Fig.~\ref{fig5}(b), the CPA and CPT
points are the extrema of this function and small deviations around these points imply a {\it quadratic sensitivity} of the two 
phenomena. 

The optimized device's  scattering properties are also verified by means of 3D FEM simulations
using an input vector of the form of Eq.~(\ref{uIs}) and following the trajectory $\phi_1=-\phi_2$ with $\phi_1\in[0,\pi]$. 
The result of the FEM simulations are shown on the $(\phi_1,\Theta)$ plane by the squares and are found to be in a good
agreement with the 1D analytical model.  The device is able to reach CPA with an input output ratio $\Theta=10^{-4}$
and a strong transmission with $\Theta=0.91$.
Let's note that in this case the high output contrast of the network can be controlled only by the relative phase of the inputs
and not the amplitudes.

\section{Discussion}

We conclude by discussing the ability of a 3-port network to exhibit a very large  output contrast (from CPA to CPT)
at the same frequency. The physical mechanism of CPA is based on the balance of the losses stemming from 
the resonator and the leakage rate of energy from the resonator to the waveguide. This is achieved by operating close to 
the resonance frequency where the losses from the HR are strong. Thus, in order to achieve CPT we need to annihilate the effect of 
losses from the HRs, and this is possible by engineering the input waves and create a destructive interference pattern at the 
positions of the HRs, which in the low frequency regime act almost like point scatterers.
In fact, the disagreement between the analytical 1D model and the FEM simulations
 in the projections in Figs.~\ref{fig5}(a) and (b) close to the CPT point, stems exactly from the fact that in the analytical model
 HRs are considered as points, while in the FEM they are small but finite. However the mechanism for CPT is the same,
as illustrated in the right panels of  Figs.~\ref{fig5} (a) and (b) for the CPT pressure profiles which acquire a minimum at the position of the resonator. 

In Section II we have found the conditions to achieve coherent perfect absorption in 
a symmetric 3-port network,  using the eigenvalues and eigenvectors of the corresponding $S$ matrix.
The conditions are directly related to the values of the transmission ($t$) and the reflection ($r$) coefficients.
It is found that the condition $r=t$ leads to an infinity of asymmetric inputs which are pefectly absorbed (asymmetric CPA)
while for $r=-2t$ a symmetric input of in-phase and equal amplitude waves from each port is perfectly absorbed (symmetric CPA).
This is in great contrast with the two port case where (for a particular device) only one input is able to achieve CPA.
The above conditions, are experimentally observed in an acoustic 3-port network composed of three connected waveguides
sideloaded by Helmlholtz resonators in the low frequency audible range.
In Section III we have studied how the output of the system is affected when the input waves deviate from the corresponding
CPA inputs. In particular we have shown that a 3-port network can exhibit both perfect absorption and perfect transmission
at the same frequency if an additional constrain is imposed. The latter dictates a value of $r=t=1/3$ for the case of an
asymmetric CPA (leading to a symmetric CPT), and  $r=-2t=2/3$ for a symmtric CPA (thus an asymmetric CPT).
The dependence of the output to input power ratio on the relative amplitude and phases of the input vectors is found analytically.
Using this expression we find that deviations the CPA and CPT are extrema of this function and thus are quadratically depended
on small amplitude and phase mismatches.
Using an optimization algorithm and the analytical expressions of the $S$ matrix elements, we further provide
particular examples of acoustic networks able to achieve both CPA and CPT (both symmetric and
asymmetric). The optimization method never converged 100$\%$ since both our model (and the realistic system) 
exhibits distributed losses and for any finite length of propagation some energy is always lost.
Finally, using FEM 3D simulation of the optimized configurations we have confirmed the transition
from perfect absorption  to a nearly perfect transmission.

The additional control over perfect absorption and the ability to go from CPA to CPT
using point scatterers, indicate that the 3-port system can be used as a unit cell to construct
complex networks with prescribed wave scattering properties. Our  theoretical results can be 
directly generalized for symmetric N-port systems, and using them build periodic networks
with different characteristics. Additionally, 3-port networks which are asymmetric, exhibit more degrees of freedom
and  could provide the  means for extra control, such as directional propagation from selected ports.

\section*{Method}

\subsection*{Determination of the elements of the scattering matrix}

\begin{figure}[th!]
\includegraphics[width=8cm]{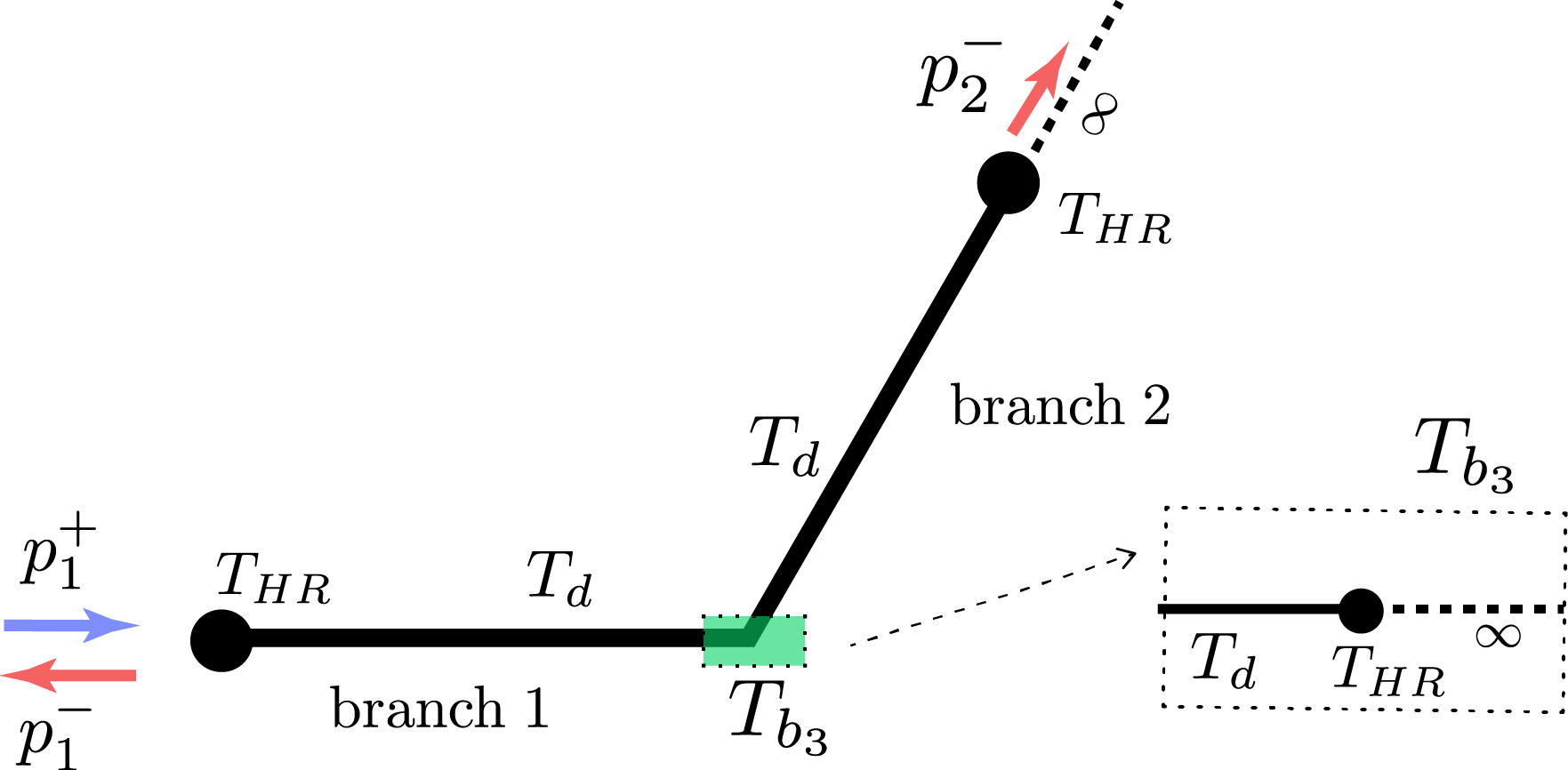}
\caption{Schematic of the 3-port system for the determination of the reflection and transmission coefficients by the Transfert Matrix Method. $T_{HR}$ describes the scattering by the side-loaded HR. $T_d$ illustrates the propagation along the waveguide of length $d$. $T_{b_3}$ describes the influence of the third branch composed by a infinite waveguide side-loaded by a HR at the distance $d$ from the connection.}
\label{appendix} 
\end{figure}

We consider the following two port scattering problem
\begin{equation} 
\label{Eq-Scattering2}
\begin{pmatrix}
p_1^- \\
p_2^- 
\end{pmatrix}
=\begin{pmatrix}
r & t \\
t & r  
\end{pmatrix}
\begin{pmatrix}
p_1^+ \\
p_2^+
\end{pmatrix} 
= S_{2\times 2}
\begin{pmatrix}
p_1^+ \\
p_2^+
\end{pmatrix} 
\end{equation}
which described the reflection and transmission from port 1 to port 2 {\it in the presence of } port 3,
as shown in Fig.~\ref{appendix}. These coefficients can then be used to describe the full 3 by 3 scattering 
problem as defined in Eq.~(\ref{Eq-Scattering}), due to the symmetry of the network.
To compute $r$ and $t$ we use the Transfer Matrix Method (TMM) which utilizes
the continuity of the pressure field and the conservation of  mass. The corresponding matrix
equation is given by
\begin{equation} 
\begin{pmatrix}
P_1 \\
U_1 
\end{pmatrix}=T
\begin{pmatrix}
P_2 \\
U_2 
\end{pmatrix}
= \begin{pmatrix}
T_{11} & T_{12} \\
T_{21} & T_{22}
\end{pmatrix}
\begin{pmatrix}
P_2 \\
U_2 
\end{pmatrix}.
\label{trm1}
\end{equation}
Here, $P_i=p_i^++p_i^-$ is the total pressure and  $U_i=S\partial_{n_i}P_i/(i\omega\rho)$ is the acoustic flux velocity 
parallel to the vector $n_i$, normal to the waveguide section at the $i-$th port. We consider sound propagation in air 
with density $\rho$ and $\omega$ denotes the angular frequency. 

Eq.~(\ref{trm1}) is obtained by considering an incident wave from port 1 and assuming no backward waves 
for the two ports remaining (no input and anechoic end) as illustrated in the Fig.~\ref{appendix}.
Then by rearranging Eq.~(\ref{trm1}) we obtain the matrix elements of Eq.~(\ref{Eq-Scattering2}) as follows. 
\begin{equation}
r = \frac{T_{11}+T_{12}/Z_c-T_{21}Z_c-T_{22}}{T_{11}+T_{12}/Z_c+T_{21}Z_c+T_{22}},\quad
t = \frac{2}{T_{11}+T_{12}/Z_c+T_{21}Z_c+T_{22}}.
\label{eq:RandT}
\end{equation}
The elements of the matrix $T= T_{HR} T_{d} T_{b_3} T_{d}  T_{HR}$  are explained in the following.
The transfer matrix $T_{HR}$ , corresponds to the HR loading the waveguide and is given by
\begin{equation} 
T_{HR}=\begin{pmatrix}
1 & 0 \\
Y_{HR} & 1
\end{pmatrix}
\end{equation}
where $Y_{HR}$ is the entrance admittance of the HR considered as a point scatterer. The expression
for the admittance can be found for example in~Ref\cite{Merkel}, where detailed information
about the viscothermal losses are also included.
$T_{d}$ is the transfer matrix for the propagation along the waveguide of length $d$ and is given by
\begin{equation} 
T_{d}=
\begin{pmatrix}
\cos(kd) & iZ_c\sin(kd) \\
i/Z_c \sin(kd) & \cos(kd)
\end{pmatrix}
\end{equation}
where $k$ is the wavenumber and $Z_c=\rho c /S$ is the characteristic impedance of the waveguide of cross-section $S$.
$T_{b_3}$ describes the influence of the third branch of the network (see Fig.\ref{appendix}), and takes the following form
\begin{equation} 
T_{b_3}=\begin{pmatrix}
1 & 0 \\
Y_{b_3} & 1
\end{pmatrix}
\end{equation}
where $Y_{b_3}$ is the entrance admittance of the branch 3 given by
\begin{equation}
Y_{b_3} = \frac{1}{Z_c} \frac{\cos (kd) (1+ Z_c Y_{HR}) + i \sin (kd)}{\cos (kd) + i Z_c \sin (kd) (1+Y_{HR})}.
\end{equation}
In our calculations we consider the effect of losses in the waveguide and in the HRs using the 
Zwikker and Kosten model\cite{Zwikker} which includes an imaginary part in the wavenumber and in the characteristic impedances of the waveguides.

\subsection*{Optimization}

From Eq. \ref{eq:RandT}, we obtain the reflexion and transmission coefficients as a function of all the system's geometrical parameters \textit{i.e.} $r_t,, \ell_n,\, r_n,\,  r_c,\, \ell_c,\, d,$ (see Fig. \ref{sketch_gen}) and of the operating frequency $f$. In order to find the device achieving both CPA and CPT, the reflexion and transmission coefficients should satisfy Eq. (\ref{cond1}) [(\ref{cond2})] with the additional constraints $r=2/3$ [$r=1/3$]. To obtain solutions satisfying this condition, a numerical optimization based on simplex derivative free method Nelder-Mead \cite{Lagarias} with 6 optimization parameters (fixing the operating frequency) under experimentally and physically reasonable constraints is used.


\section*{Acknowledgements}

This work is supported by "Le Mans Acoustique" and funded by the "Pays de la Loire" French Region " and  the "European Regional Development Fund". I. Brouzos appreciates financial support from IKY (State Scholarships Foundation).

\section*{Author contributions}

O.R., V.A. and G.T. performed the theoretical simulations; I.B. performed the optimization process. O.R. and V.A. performed the experiments; O.R., V.A., G.T. and I.B. wrote the manuscript. All authors contributed to data analysis and discussions.

\section*{Additional Information}
Competing Interests:  the authors declare that they have no competing interests.

\end{document}